\documentclass[%
 reprint,
 amsmath,amssymb,
 aps,
]{revtex4-2}

\usepackage{graphicx}
\usepackage{dcolumn}
\usepackage{bm}

\begin{document}

\preprint{APS/123-QED}

\title{A possibility of describing a sequential gauge symmetry as a part of a tetrahedron rotational symmetry and its implications}

\author{\small Jae Jun Kim}
 
\affiliation{
 South Carolina Department of Education\\
 Columbia, South Carolina\\
The United States
 \\
}%


\date{\today}

\begin{abstract}
We present a method in which the dimension-five and -six terms in a sequential gauge can coexist with that in a non-sequential gauge under a tetrahedron rotational symmetry.  We start with the contents in the model proposed in a sequential gauge, meeting the conditions for the contents to be described in the rotational symmetry.  Given that the breaking of the symmetry takes place in the dimension-five and -six terms via the flavor-specific Higgs scalars, we address how introducing the non-sequential contents can be realized.  In doing so, we illustrate that the terms in the sequential gauge could be understood as the higher-order terms, the correction terms, and that in the non-sequential gauge as the leading order term.  
Our result could be understood as an illustration of introducing the sequential gauge as a reason for the mixing as having non-zero left-handed rotational matrix in the quark and the lepton sectors, how the mixing in the charged lepton sector can be small compared to that of the neutral sector, and how the sequential and the non-sequential gauge terms can coexist without introducing super-symmetry.
\end{abstract}

\maketitle


\section{Introduction}	
\label{intro}
Constructing and testing a model under the non-abelian flavor symmetries has been successful.  Many studies have been conducted to enhance our understanding. 
\cite{reference1, reference17, reference18, reference19, reference20, reference21}.  

Among them, accommodating the sequential gauge was crucial since the generation of the mass in the charged and the neutral fermion sectors could be realized within a model while securing the nature of the anomaly-free.  The sequential gauge symmetry in previous literature was about realizing the generation of the mass matrix even when the breaking of the global $U(1)$ takes place by assigning different charges to that of the different generation \cite{reference4, reference7}.

In this study, to understand the nature of the mass matrices and the mixing further, we explore a possibility of accommodating the contents in the fermion sectors introduced in the sequential $U(1)$ gauge as correction terms, describing them with that of the non-sequential terms under $A_4$, a tetrahedron rotational symmetry.  Introducing the correction terms in the context of a four-dimensional super-symmetry under $A_4$ had shown success \cite{reference10,reference17}, but we try to do so without it.  For that, we start with the contents of the model studied in \cite{reference4}, where the sequential structure in the contents are realized, securing the nature of the anomaly-free \cite{reference5, reference8}.  After that, we extend it by adding non-sequential fields.  

In essence, we tried to seek the answers to the following questions: Can we realize the structure of the mass matrix under a sequential gauge, but taking a rotational symmetry such as $A_4$ into account, without introducing super-symmetry?  Can the sequential and the non-sequential invariant terms coexist within a model and what is an advantage of doing so?  

In section 2, we describe the contents of the model.  In section 3, we describe how the model works out in the quark sector.  In section 4, we do the same as that in section 3 but for the lepton sector.  In section 5, we describe conditions needed to realize the mass matrix given all the contents.  In section 6, we describe how our model can accommodate the empirical mixing parameters.  In section 7, we discuss our result.

\section{Contents}	
\label{contents}
We start with the sequential part and will add the non-sequential fields later.  The entire contents introduced in our model are presented in the Appendix.  

First, we assign the charges to the contents of the model studied in \cite{reference4} as they were for all the fermions.  However, given that the Higgs scalars in $A_4$ do not have the sequential structure in the study, we need to add the charges and the representation under $A_4$ for the scalars.  
The considered symmetry is,
\begin{equation}
G = SU(2)_L \times U(1)_Y \times U(1)_X \times A_4 {,}
\end{equation}
where $U(1)_X$ is auxiliary charge and $A_4$ is a tetrahedron rotational symmetry.
{The field and scalar} contents of the model for the sequential part are given by:
\begin{equation}
\begin{split}
&\psi_{iL} \sim \left[ \hspace{0.4mm} 2, {-1}, 4n, 1_i \hspace{0.4mm} \right],\\
&\phi_{SM} \sim \left[ \hspace{0.4mm} 2, {1}, c_1, 3\hspace{0.4mm} \right]\\
&\phi_{L} \sim \left[ \hspace{0.4mm} 1, 0, c_2, 3\hspace{0.4mm} \right]
,\phi_{Q} \sim \left[ \hspace{0.4mm} 1, 0, c_3, 3\hspace{0.4mm} \right],\\
&l_{i,R} \sim \left[ \hspace{0.4mm} 1, {-2}, 5n, 1_i \hspace{0.4mm} \right]\label{particlecontent}\\
&u_L, d_L \sim \left[ \hspace{0.4mm} 2, 1/3, 0, 3\hspace{0.4mm} \right],\\
&u_{i,R} \sim \left[ \hspace{0.4mm} 1, 4/3, -3n, 1_i \hspace{0.4mm} \right],\\
&d_{i,R} \sim \left[ \hspace{0.4mm} 1, -2/3, 3n, 1_i \hspace{0.4mm} \right],\\
&\phi_{NU} \sim \left[ \hspace{0.4mm} 1, 0, c_1, 3\hspace{0.4mm} \right]
\end{split}
\end{equation}
, where {$1_i= [\hspace{0.4mm} 1,1',1'' \hspace{0.4mm}]$, representing the generation, $1'$ and $1''$ in $A_4$, to indicate the non-trivial and $1$ the trivial representation} respectively.

All the charges for the charged fermions for $SU(2)_L$, $U(1)_Y$ and $U(1)_X$ are introduced in \cite{reference4}, but the representations in $A_4$.  Taking the $U(1)_X$ charges defined as in \cite{reference3}, the charges in our study can be written as a function of,
\begin{equation}
n = \frac{3}{4}-\frac{i}{4}
\end{equation}
, where $i = 1,2,3$.  Such convention is introduced to take the assignment of the charges differently for different generation of families \cite{reference4}.  With that, $U(1)_X$ charge for the Higgs scalars can be assigned as,
\begin{equation}
c_1 = 0, \hspace{0.4mm} c_2 = \frac{1}{4}, \hspace{0.4mm} c_3 = \frac{3}{4}
\end{equation}
.  In the contents, Equation 2, $\psi_{L,R}$ and $l_{R}$ represent left-handed and right-handed leptons.  $\phi_{SM}$ represents a Higgs scalar similar to that in the standard model but ours in $A_4$ triplet, with a zero $U(1)_X$, which is going to be held under $Z_3$.  
As for the flavor-specific scalars, $\phi_{L}$ is for the lepton sector and $\phi_{Q}$ for the quark sector, $u_{L}$ and $d_{L}$ are left-handed quark fields and $u_{i,R}$ and $d_{i,R}$ being right-handed in each family generation.  
We also introduce $\phi_{NU}$, one more Higgs scalar, for the neutrino sector \cite{reference3}.  All the fermion fields without superscript can be assumed to be sequential terms.  

The basis we choose for the study is where the generator,
\begin{equation}
S=
\begin{array}{c|ccccc|c}
&& 1 & 0 & 0 && \\
&& 0 & -1 & 0 &&\\
&& 0 & 0 & -1&& \\
\end{array}
\end{equation}
, is diagonal \cite{reference17}.


Throughout our study, we adopt the following naming convention to differentiate mass and rotational matrices in different sectors,
$UQ, U$: Up-quark, $DQ$: Down-quark, $CL, L$: Charged lepton, $DN$: Dirac neutrino, $MN$: Majorana neutrino, $D$: Diagonal, $SE$: Sequential, $NS$: Non-sequential $NU$: Neutrino.

\section{Quark sector}	
\label{firstsector}
With all the sequential contents, starting with the up-side quark, our Yukawa interaction under $A_4$ for the sequential part can be written as,
\begin{equation}
\begin{split}
-&L^{SE}_{U} =h_1 {\overline{u}}_{L}\overline{\phi}_{SM} u_{1R} \phi^{2}_{Q}\Lambda^{-2} +
h_2 {\overline{u}}_{L} \overline{\phi}_{SM} u_{2R} \phi_{Q} \Lambda^{-1} \\
& + h_3 {\overline{u}}_{L}\overline{\phi}_{SM} u_{3R} + h.c.
\end{split}
\end{equation}
, where $h_{i}$ represents the Yukawa interaction scale, and $SE$ stands for the sequential part in our study.  Equation 6 has a similar structure as that in \cite{reference1}, but have different dimensions in the first and the second terms due to $c_i$ having different charges for different generations of family.

To realize the mass matrix, we add a $Z_3$ discrete symmetry as a part of the content to prevent the terms where the coupling of $\phi_{L}$ and $\phi_{Q}$, and that of $u_L$ and $d_{i,R}$, and $d_L$ and $u_{i,R}$ could take in place.  \cite{reference18} also has $Z_3$ discrete symmetry as a part of their model, although the assigned charges are different from that in our study.  We assign the discrete charge for the charged lepton fields as,
\begin{equation}
\begin{split}
\psi^{SE}_{1,L} \sim \omega, \psi^{SE}_{2,L} \sim \omega^2, \psi^{SE}_{3,L} \sim \omega,
v^{SE}_{1,R} \sim \omega^2, v^{SE}_{2,R} \sim \omega, v^{SE}_{3,R} \sim \omega^2
\end{split}
\end{equation}
.  The charge for the flavor-specific scalar and the lepton fields as,
\begin{equation}
\begin{split}
&\phi^{}_{Q} \sim \omega, \phi^{}_{L} \sim \omega, \phi^{}_{NU} \sim \omega,\\ &\psi_{L} \sim \omega, l_{i,R} \sim \omega
\end{split}
\end{equation}
, where $i = 1, 2, 3$.

On the same token, with all the assignment of charges, the Yukawa interaction for the down quark can be written as,
\begin{equation}
\begin{split}
-&L^{SE}_{D} =h_4 {\overline{d}}_{L}\phi_{SM} d_{1R} \phi^{2}_{Q} \Lambda^{-2} +
h_5 {\overline{d}}_{L}\phi_{SM} d_{2R} \phi_{Q} \Lambda^{-1}\\
& + h_6 {\overline{d}}_{L}\phi_{SM} d_{3R} + h.c.
\end{split}
\end{equation}
.  One of the differences between our study and that in \cite{reference1} and \cite{reference3} is that each generation of the family has one leading order term in each column, with different dimension.  Such is due to our constructing the Lagrangian under $A_4$, and having the three Higgs scalars with different $U(1)_X$ charges.

The Yukawa interaction for the non-sequential terms can be written as,
\begin{equation}
\begin{split}
-&L^{NS}_{D} =h^{}_{40} {\overline{d^{}}_{L}}\phi_{SM} d^{NS}_{1R} +
h^{}_{50} {\overline{d}}_{L}\phi_{SM} d^{NS}_{2R} \\
& + h^{}_{60} {\overline{d}}_{L}\phi_{SM} d^{NS}_{3R} + h.c.
\end{split}
\end{equation}
, where $h^{}_{40,50,60}$ corresponds to the scale for the non-sequential terms.  The non-sequential contents are listed in the Appendix.  

Taking the Yukawa interaction into account, 
we can break $A_4$ to $Z_3$.  For that, we assign the vacuum expectation value, v.e.v, for the Higgs scalars as,
\begin{equation}
\begin{split}
&\phi_{SM} = [\phi_{SM1},\phi_{SM2},\phi_{SM3}] = [v_1,v_2,v_3] \\
\end{split}
\end{equation}
, where,
\begin{equation}
\begin{split}
v_1 \sim v_2 \sim v_3 \sim v_S
\end{split}
\end{equation}
in order to diagonalize of the mass matrix.  
The same can be realized with $\phi_{L}$ and $\phi_{Q}$ although $Z_3$ needs to be broken.  For the quark sector,
\begin{equation}
\begin{split}
&\phi_{Q} = [\phi_{Q1},\phi_{Q2},\phi_{Q3}] = [v_{Q1},v_{Q2},v_{Q3}] \\
\end{split}
\end{equation}
.  For the lepton sector,
\begin{equation}
\begin{split}
&\phi_{L} = [\phi_{L1},\phi_{L2},\phi_{L3}] = [v_{L1},v_{L2},v_{L3}] \\
\end{split}
\end{equation}
.  With all the assignment, we can break $A_4$ down by,
\begin{equation}
\begin{split}
\phi_{L,i} = [ \hspace{0.3mm} 0, v_{i}, 0 \hspace{0.3mm} ]^T
\end{split}
\end{equation}
, where $i = 1,2,3$, and v.e.v of the flavor-specific part is assumed to be in a similar order for both the quark and the lepton sector, although they could be in a different scale \cite{reference6}.  Using the product rule of $A_4$ in \cite{reference9}, we end up with the following mass matrix for the up-quark sector,
%
\begin{equation}
M^{SE}_U =
\begin{array}{c|ccccc|c}
&& h_1v_S v^2 & h_2v_S v & h_3v_S &&\\
&& h_1v_S v^2 & h_2v_S v \omega & h_3v_S \omega^2 &&\\
&& h_1v_S v^2 & h_2v_S v \omega^2 & h_3v_S \omega &&\\
\end{array}
\end{equation}
, where we assume the contents $v^2$ in the first column, dimension-six, are scaled by $\Lambda^{-2}$ and that $v$ of the second column, dimension-five, is scaled by $\Lambda^{-1}$, throughout our study of the sequential part.  By including the leading order terms only thus the dimension-six terms in the last column, a term such as,
\begin{equation}
{\overline{u}}_{L}\overline{\phi}_{SM} u_{1R} \phi^{*}_{SM}\phi_{SM} \Lambda^{-2}
\end{equation}
, is not introduced as a part of the third-family.  



Once we end up with the matrix after the flavor symmetry breaking, even with the sequential structure, what we have shown is that we could end up with a diagonalized matrix.  We can diagonalize the matrix by doing rotations with \cite{reference1},
\begin{equation}
\sqrt{3} \cdot U_L =
\begin{array}{c|ccccc|c}
&&1 & 1 & 1 && \\
&& 1 & \omega & \omega^2 && \\
&&1 & \omega^2 & \omega && \\
\end{array}
, U_R = 1
\end{equation}
.  With that, we end up with,
\begin{equation}
U^{+}_{L} M^{SE}_U U^{}_{R}=
\begin{array}{c|ccccc|c}
&& M^{d}_{1} & ... & ... &&\\
&& ... & M^{d}_{2} & ... &&\\
&& ... & ... & M^{d}_{3} &&\\
\end{array}
\end{equation}
.  Interestingly, following \cite{reference4}, the non-diagonal terms in the mass matrix is going to be only in the first two columns in the quark sector, the dimension-six and -five columns, in the leading order.

For instance, when we break the symmetry for $\phi_Q$, we have,
%
\begin{equation}
M^{'SE}_U =
\begin{array}{c|ccccc|c}
&&h_1v_S v^2 & h_2v_S v & h_3v_S &&\\
&& h_1v_S v'^2 & h_2v_S v' \omega & h_3v_S \omega^2 &&\\
&& h_1v_S v''^2 & h_2v_S v'' \omega^2 & h_3v_S \omega &&\\
\end{array}
\end{equation}
.  It happened to be so since the $U(1)_X$ charge is asymmetrically assigned in the sequential gauge model.  By doing so, we do not have the non-diagonal terms in the last column, the third column, of the matrix.

However, we can end up with Equation 20 only if we break the symmetry by assigning different values of v.e.v, to the quark-specific Higgs, $\phi_{Q}$, which is associated with the first and the second generation of family only in the case of the quark sector.  Adjusting the numerical value for the breaking can be realized to meet the size of the mixing that is measured in empirical studies.

In other words, while we do with breaking $\phi_{Q}$, we cannot do the same with $\phi_{SM}$, to realize the structure of the mass matrix as shown in \cite{reference4},
\begin{equation}
U^{+}_{L} M^{'SE}_U U^{}_{R}=
\begin{array}{c|ccccc|c}
&& V^{i}_{11} & ... & ... &&\\
&& V^{i}_{21} & V^{i}_{22} & ... &&\\
&& V^{i}_{31} & V^{i}_{32} & V^{i}_{33} &&\\
\end{array}
\end{equation}
, where the breaking takes in place in the first two columns only in the mass matrix.  

Note that $i$ as a superscript in Equation 21 and other mass and mixing matrices later on is written to distinguish matrices in different sectors, not different generations.  In this case, it is up-side quark sector so $i = UQ$.

With the definition of,
\begin{equation}
\Delta M^{SE} = M^{'SE}_{U}-M^{SE}_{U}
\end{equation}
.  Then the correction terms, due to the breaking of $Z_3$, can be written as,
\begin{equation}
\Delta M=
\begin{array}{c|ccccc|c}
&& ... & ... & ... &&\\
&& M^{i}_{21} & M^{i}_{22} & M^{i}_{23} &&\\
&& M^{i}_{31} & M^{i}_{32} & M^{i}_{33} &&\\
\end{array}
\end{equation}
, which happened to be the structure of the mass matrix representing the mediation between different sectors \cite{reference3}.

With that, we could end up with non-zero terms in the mass matrix after the rotation in the third family.  In order not to have such, we need the correction terms to be in a form of,
\begin{equation}
\Delta M^{SE}=
\begin{array}{c|ccccc|c}
&& ... & ... & ... &&\\
&& M^{i}_{21} & M^{i}_{22} & ... &&\\
&& M^{i}_{31} & M^{i}_{32} & ... &&\\
\end{array}
\end{equation}
, where we see all zeros in the third column and that can be realized by breaking $\phi_{Q}$ in the quark sector, while having $\phi_{SM}$ intact under $Z_3$.  Having such corrections could be understood as the mediation from the non-sequential sector to the sequential sector after the breaking of $\phi_{Q}$ and vice versa.  It could be considered as a sum of the mass terms in the sequential and the non-sequential sectors too.



With all that, $\phi_{SM}$ can stay unbroken under $Z_3$ and it has an implication: We can extend the model to have more contents from the non-sequential part that was described in \cite{reference3}, while keeping all the sequential terms as they are.

In other words, we can introduce more contents from that in the non-sequential gauge, including that in \cite{reference3}, while keeping all the sequential contents in Equation 2 to build up correctional terms.  We are able to add their charged fermion contents, $\psi_{L}$ and $l_{R}$ with a same $U(1)_X$ charge for all the family.  
We add the following to be a part of the content of the model,
\begin{equation}
\begin{split}
&\psi^{NS}_{L} \sim \left[ \hspace{0.4mm} 2, -1, 0, 3 \hspace{0.4mm} \right],\\
&{l^{NS}_{i,R}} \sim \left[ \hspace{0.4mm} 1, -2, 0, 1_i \hspace{0.4mm} \right],\\
\end{split}
\end{equation}
. For the quark sector, we add,
\begin{equation}
\begin{split}
&u^{NS}_{i,R} \sim \left[ \hspace{0.4mm} 1, 4/3, 0, 1_i\hspace{0.4mm} \right],\\
&d^{NS}_{i,R} \sim \left[ \hspace{0.4mm} 1, -2/3, 0, 1_i\hspace{0.4mm} \right],\\
\end{split}
\end{equation}
, where $i = 1,2,3$ and $1_i$ represents singlet in $A_4$. to that we listed in Equation 2.  $NS$ stands for that of the non-sequential.  The fields in Equation 25 and 26 were introduced in \cite{reference3} before.

With all that, the Yukawa interaction in the up quark sector is going to have more components, sequential and non-sequential parts together.  Taking the sequential-terms in Equation 6, the interaction in the up side of the quark sector can be written as,
\begin{equation}
\begin{split}
&L_U =  L^{SE}_U + h_7 {\overline{u}}_{L}\phi_{SM} u^{NS}_{1R} +
h_8 {\overline{u}}_{L}\phi_{SM} u^{NS}_{2R} +\\
&h_9 {\overline{u}}_{L}\phi_{SM} u^{NS}_{3R} + h.c.
\end{split}
\end{equation}
, where $L^{SE}$ is the sequential terms in Equation 6 and the rest is what was introduced before as the non-sequential terms in \cite{reference3}.

Writing it in a different manner, for the quark sector, having both the sequential and the non-sequential parts, expanding them and writing in terms of the mass matrix after the breaking, the flavor mass Lagrangian can be written as,
\begin{equation}
L^{}_U = \overline{u}^{m}_L M^{NS}_U u^{NS,m}_{R} + \overline{u}^{m}_{L} M^{SE}_U u^{SE,m}_{R} + h.c
\end{equation}
, where,
\begin{equation}
\begin{split}
{u}^{m}_L = [u^{m}_{1L},u^{m}_{2L},u^{m}_{3L}],
{u}^{i,m}_R = [u^{i,m}_{1R},u^{i,m}_{2R},u^{i,m}_{3R}]^{T}
\end{split}
\end{equation}
, where $m$ indicates the content in the mass Lagrangian after the breaking of the symmetry and $i$ is either $NS$ or $SE$.  

When the mass matrix for the sequential and the non-sequential part are added, we end up with, 
\begin{equation}
M_U =
\begin{array}{c|ccccc|c}
&&h_7v_S & h_8v_S & h_{9}v_S &&\\
&& h_7v_S + h_1v_S v'^2 & h_8v_S \omega + h_2v_S v' \omega & h_{9}v_S \omega^2 &&\\
&& h_7v_S + h_1v_S v''^2 & h_8v_S \omega^2 + h_2v_S v'' \omega^2 & h_{9}v_S \omega &&\\
\end{array}
\end{equation}
.  The same can be realized on the down side of the quark sector too.  With the rotation of the mass matrix by $U_L$ and $U_R$, we can realize the mass matrix as in Equation 20, which is where the sequential terms could be understood as the correction terms. 





\section{Lepton sector}	
\label{sectorsecond}
In the quark sector, the model that leads us to accommodate the non-sequential and the sequential gauge terms could be an indication of the breaking of the family universality. 
Furthermore, our being able to explain the sequential structure as not breaking $\phi_{SM}$ indicates the opportunity to describe the sequential terms as the correction terms.  



However, we have a different scenario in the charged lepton sector.  
The sequential structure of assigning the charges in the sector is in essence same as that in the quark sector.  A difference is that $U(1)_X$ charges for the left-handed leptons do have the sequential structures but that in the quark sector, we do not.  Those in the right-handed are different in both the lepton and the quark sector.  In other words, they both,
\begin{equation}
\begin{split}
\psi_{i,L} \sim \left[ \hspace{0.4mm} 2, {-1}, 4n, 1_i \hspace{0.4mm} \right],
{l_{i,R}} \sim \left[ \hspace{0.4mm} 1, {-2}, 5n, 1_i \hspace{0.4mm} \right]
\end{split}
\end{equation}
, are $A_4$ singlets.

For that reason, the way we introduced the sequential terms for the quark sector cannot be realized in the same manner.  The left-handed leptons need to be under a singlet in $A_4$ instead of having all three in a triplet.  The dimension-six and -five terms can be written in a similar manner, but a term such as,
\begin{equation}
\begin{split}
h^{}_{12}{\overline{\psi}^{}}^{}_{3L} \phi^{}_{SM} \psi^{}_{3R}
\end{split}
\end{equation}
, cannot be introduced as part of the Yukawa interaction due to their not having an invariant term.

The only ones that can be introduced are the terms where the non-sequential left-handed fields coupled with $\phi_{SM}$, which is the non-sequential part, and a sequential part of right-handed field.  With that, we have,
\begin{equation}
\begin{split}
-&L^{SE}_L = h^{}_{10} {\overline{\psi}}^{}_{1L}\phi^{}_{SM} {l_{1R}} \phi^{*2}_{L} \Lambda^{-2} +
h^{}_{11} {\overline{\psi}}^{}_{2L}\phi^{}_{SM} {l_{2R}} \phi^*_{L} \Lambda^{-1}\\
&h^{}_{12} {\overline{\psi}^{}}^{}_{L} \phi^{}_{SM} {l^{}_{3R}} + h.c.
\end{split}
\end{equation}
, where the last term came from the non-sequential part of the left-handed field.  

It happens to be the case since $u_L$ and $d_L$ have their $U(1)_X$ charge being same for different generations, whereas all other sectors require some degree of mediation from the non-sequential parts.  

That being said, just like that in the quark sector, we could realize the same patterns of the mass matrix as in \cite{reference4} by breaking the $Z_3$ symmetry in $\phi_{L}$.  For the non-sequential part of the charged lepton sector, then we end up with,
\begin{equation}
M^{SE}_L =
\begin{array}{c|ccccc|c}
&&h^{}_{10}v^{}_S v^2 & h^{}_{11}v^{}_S v^{}_{} & h^{}_{12}v^{}_S &&\\
&& h^{}_{10}v^{}_S v'^2 & h^{}_{11}v^{}_S v'_{} \omega & h^{}_{12}v^{}_S \omega^2 &&\\
&& h^{}_{10}v^{}_S v''^2 & h^{}_{11}v^{}_S v^{''}_{} \omega^2 & h^{}_{12}v^{}_{S} \omega &&\\
\end{array}
\end{equation}
, where $v = v_L$.  So we do need a mediation from the sequential sector to the non-sequential sector, the third column in Equation 34, to realize the matrix.

By not breaking $Z_3$ of $\phi_{SM}$, the correction term can be absorbed as a part of the non-sequential term thus the matrix has a same structure as in Equation 30, from which the mass matrix in \cite{reference4} after rotation,
\begin{equation}
U^{+}_{L}M^{SE}_{L}U^{}_R=
\begin{array}{c|ccccc|c}
&& V^{i}_{11} & V^{i}_{12} & ... &&\\
&& V^{i}_{21} & V^{i}_{22} & ... &&\\
&& ... & ... & V^{i}_{33}&&\\
\end{array}
\end{equation}
, where $i=CL$, charged lepton sector, can be realized.  



Note that the structure of breaking as adjusting $v'$ and $v''$ in the lepton sector, which is shown in Equation 34, is different from Equation 20.  Having such can be understood as an indication that we need at least two flavor-specific Higgs scalars, $\phi_Q$ and $\phi_L$.

As for the neutrino sector, just like that in the quark sector, we introduce more contents for the non-sequential part as described in \cite{reference3}, 
\begin{equation}
\begin{split}
&v^{NS}_R \sim \left[ \hspace{0.4mm} 1, 0, 0, 3\hspace{0.4mm} \right],\\
&\phi_{NS} \sim \left[ \hspace{0.4mm} 2, -1, 0, 1\hspace{0.4mm} \right]
\end{split}
\end{equation}
, where the first term is right-handed neutrino triplet in $A_4$ and the second one being a Higgs scalar singlet.  With the two being added to the content, we can realize the non-sequential part of the neutrino mass matrix described in \cite{reference3}, where the dimension-five operator can be used to describe the type 1 seesaw mechanism.

As for the sequential part, we also need to introduce,
\begin{equation}
\begin{split}
&v^{SE}_{i,R} \sim \left[ \hspace{0.4mm} 3, 0, n, 1_i\hspace{0.4mm} \right]
\end{split}
\end{equation}
, by which the type 3 seesaw mechanism \cite{reference4} can be described with all the sequential part of contents and $\phi_{NU}$.  Note that Equation 37 is the only triplet in $SU(2)_L$ in the model.


With all that, the Yukawa interaction for the non-sequential part under $A_4$ is same as that was described in other literature including \cite{reference3},
\begin{equation}
\begin{split}
-L^{NS}_{NU} = h_{13} {\overline{\psi}}^{}_{L} v^{}_{R} \phi^{}_{NS} +
h_{14} \overline{v}^{}_{R} v^{c}_{R} \phi^{}_{NU} +
m\hspace{0.6mm}\overline{v}^{}_{R} v^c_R + h.c.
\end{split}
\end{equation}
, where the first one is for the Dirac, the second is for the Majorana, and the third one is the Majorana bare mass term.  


Taking the contents and the $Z_3$ charge as in Equation 7, 8 and 9, we build the Yukawa interaction for the sequential part.  For the Dirac neutrino sector, we have,
\begin{equation}
\begin{split}
-&L^{SE}_{NU,D} =
h_{15} {\overline{\psi}}^{}_{1L} v^{SE}_{1R} \phi^{}_{SM} \phi^{}_{Q}{\phi}^{}_{Q} \Lambda^{-2} +\\
&h_{16} {\overline{\psi}}^{}_{2L} v^{SE}_{2R} \phi^{}_{SM} {\phi}^{}_{Q} \Lambda^{-1} +\\
&h_{17} {\overline{\psi}}^{}_{3L} v^{SE}_{3R} \phi^{}_{NS} \phi^{}_{NU} \phi^{}_{NU} \Lambda^{-2} + h.c.\\
\end{split}
\end{equation}
, where, interestingly, the terms with $\phi_L$ is not a part of the expression due to the presence of $Z_3$ discrete symmetry.  Given the Lagrangian, the mass matrix shown in \cite{reference4},
\begin{equation}
U^{+}_{L}M^{SE}_{DN}U^{}_R=
\begin{array}{c|ccccc|c}
&& V^{i}_{11} & ... & ... &&\\
&& V^{i}_{21} & V^{i}_{22} & V^{i}_{23} &&\\
&& V^{i}_{31} & V^{i}_{32} & V^{i}_{33}&&\\
\end{array}
\end{equation}
, where $i=DN$, can be realized by breaking $Z_3$ in $\phi_{Q}$ in each generation with different $h_i$.  

Note also that the last term,
\begin{equation}
\begin{split}
h_{17} {\overline{\psi}}^{}_{3L} v^{SE}_{3R} \phi^{}_{NS} \phi^{}_{NU} \phi^{}_{NU} \Lambda^{-2}
\end{split}
\end{equation}
, a leading-order term in the sequential part, in Equation 39 are in dimension-six.
Having such could be understood as a reason for having the dimension-six term, for $M_{23}$, in the last column in the Dirac neutrino mass matrix \cite{reference4}.  

Given the sequential term in the last column is in dimension-six, we do need leading-order terms in both the sequential and the non-sequential parts to realize the mass matrix in our model, just like that for the charged lepton sector.

In the the Majorana neutrino sector, we have the Yukawa interaction as,
\begin{equation}
\begin{split}
-&L^{SE}_{MN} = h^i_{18} {\overline{v}}^{SE}_{i,R} v^{SE}_{i,R} \phi^*_{Q} \phi^{}_{Q} \Lambda^{-1} + \\
&h^i_{19} {\overline{v}}^{SE}_{i,R} v^{SE}_{i,R} \phi^*_{L} \phi^{}_{L} \Lambda^{-1}+\\
&h^i_{20} {\overline{v}}^{SE}_{i,R} v^{SE}_{i,R} \phi^*_{SM} \phi^{}_{SM} \Lambda^{-1} +\\
&h^i_{21} {\overline{v}}^{SE}_{i,R} v^{SE}_{i,R} \phi^*_{NU} \phi^{}_{NU} \Lambda^{-1} +\\
&h^{}_{22} {\overline{v}}^{SE}_{3,R} v^{SE}_{1,R} \phi^{*}_{Q} \phi^{}_{L} \Lambda^{-1} +\\
&h^{}_{23} {\overline{v}}^{SE}_{1,R} v^{SE}_{3,R} \phi^{}_{Q} \phi^{*}_{L} \Lambda^{-1} +\\
&h^{}_{24} {\overline{v}}^{SE}_{2,R} v^{c,SE}_{2,R} \phi^{}_{Q} \phi^{*}_{L} \Lambda^{-1} + h.c\\
\end{split}
\end{equation}
, where $i=1,2,3$. 
All the terms are in dimension-five, same as in \cite{reference4}.  The sequential fields come in pairs with its own complex conjugate, since that is only one triplet in $SU(2)_L$.



For that reason, the Majorana mass matrix,
\begin{equation}
U^{+}_{L}M^{SE}_{MN}U^{}_R=
\begin{array}{c|ccccc|c}
&& V^{i}_{11} & ... & V^{i}_{13} &&\\
&& ... & V^{i}_{22} & ... &&\\
&& V^{i}_{31} & .. & V^{i}_{33}&&\\
\end{array}
\end{equation}
, where $i=MN$, Majorana neutrino, cannot be realized due to Equation 42 being invariant from breaking $Z_3$ thus we need the terms from the non-sequential sector.  


In summary, the breaking of the family universality could be taken as the non-leading-order terms and could be understood as a reason for having the mixing under a rotational symmetry.  At the same time, we do need the contents from both the non-sequential and the sequential part to realize the mass matrices, with $Z_3$ discrete symmetry being introduced to constrain some invariant terms from appearing.

\section{Understanding the mixing in each sector and its implications}
\label{analysis}
We describe the patterns of mixing among the constituents in the sequential parts of the mass matrices and why we need non-sequential part as a part of the model.


We start with the quark sector.  As described in the previous sections, the mass matrix of the sequential terms can be written as,
\begin{equation}
M^{SE}_{Q,L}=
\begin{array}{c|ccccc|c}
&& ... & ... & ... &&\\
&& M^{i}_{21} & M^{i}_{22} & ... &&\\
&& M^{i}_{31} & M^{i}_{32} & ... &&\\
\end{array}
\end{equation}
, where $M_i$ indicates the size of the elements in the mass matrix representing each correctional term.  

After rotating the matrix by $U_L$, we end up with,
\begin{equation}
U^{+}_L M^{SE}_Q =
\begin{array}{c|ccccc|c}
&& V^{i}_{11} & ... & ... &&\\
&& V^{i}_{21} & V^{i}_{22} & ... &&\\
&& V^{i}_{31} & V^{i}_{32} & V^{i}_{33} &&\\
\end{array}
\end{equation}
, where $V$ is to indicate the size of the elements after the rotation.  The size of the $V^i$ can be determined from the empirical measurement.  Given the structure, we can constrain the elements where we have zeros in the leading order, which is $...$ in the matrix.  For instance, the elements $V^i_{12}$ needs to be zero in the leading order in the quark sector and that is the only one with zero since the breaking in third column does not take in place.  From the previous two equations,  then we have,
\begin{equation}
V^{i}_{12} = M^{i}_{22} + M^{i}_{32} = 0
\end{equation}
.  Note that $M_{22}$ and $M_{32}$ have the same Yukawa scale, which in this case, $h_2$ in Equation 7.  
Given all that, we can write,
\begin{equation}
M^{i}_{22} = h^{}_2 v^{}_{SM} v'_{Q} \omega, M_{32} = h^{}_2 v^{}_{SM} v''_{Q} \omega^2
\end{equation}
, from which we end up with,
\begin{equation}
v'_{Q} + v''_{Q} \omega  = 0
\end{equation}
.  As illustrated in Equation 45, we do not see mixing on the dimension-four sequential part so $V_{33}$ can be determined from that of the non-sequential terms.

Note that, in this section, $v'_Q$ is not the size of v.e.v after breaking but the difference between the size of v.e.v before and after breaking.


As for the charged lepton sector, the matrix after the rotation can be written as,
\begin{equation}
U^{+}_L M^{SE}_L =
\begin{array}{c|ccccc|c}
&& V^{i}_{11} & V^{i}_{12} & ... &&\\
&& V^{i}_{21} & V^{i}_{22} & ... &&\\
&& ... & ... & V^{i}_{33} &&\\
\end{array}
\end{equation}
.  Given $V^{i}_{31} = V^{i}_{32} = 0$ in the first two columns in the leading order, we have,
\begin{equation}
\begin{split}
&V^{i}_{31} = \omega M^{i}_{21} + \omega^2 M^{i}_{31} = 0,\\
&V^{i}_{32} = \omega M^{i}_{22} + \omega^2 M^{i}_{32} = 0
\end{split}
\end{equation}
.  Taking the first expression in Equation 50, in the linear order, we end up with,
\begin{equation}
\begin{split}
v'^{}_L + \omega^2 v''^{}_L \sim 0
\end{split}
\end{equation}
, where the terms in the order of $v'^2$ are not included.  On the other hand, $V_{32} = 0$ leads us to, 
\begin{equation}
\begin{split}
\omega^2 v'^{}_{L} + v''^{}_{L} = 0
\end{split}
\end{equation}
.  Given Equation 51 and 52, we do not see the mixing in the charged lepton sector in the leading order to realize the structure of the mass matrix.  Even in the second order, having a mixing leads us to $v \sim v'$, which is not realistic since $v \gg v'$.  For that reason, Equation 51 and 52 could be understood as a reason for
the mixing to be small in the charged lepton sector.

As for the neutrino sector, the mass matrix before the rotation by $U_L$ and $U_R$ can be written as,
\begin{equation}
M^{SE}_{NU,D}=
\begin{array}{c|ccccc|c}
&& ... & ... & ... &&\\
&& M^{i}_{21} & M^{i}_{22} & M^{i}_{23} &&\\
&& M^{i}_{31} & M^{i}_{32} & M^{i}_{33} &&\\
\end{array}
\end{equation}
.  For the Dirac neutrino part, we need to have the following after the rotation,
\begin{equation}
U^{+}_{L} M^{SE}_{NU,D} =
\begin{array}{c|ccccc|c}
&& V^{i}_{11} & ... & ... &&\\
&& V^{i}_{21} & V^{i}_{22} & V^{i}_{23} &&\\
&& V^{i}_{31} & V^{i}_{32} & V^{i}_{33} &&\\
\end{array}
\end{equation}
, where $V^{i}_{12} =0$ leads us to a similar expression as in Equation 47 and 48 in the quark sector.  

With that, $V^{i}_{12}$ in Equation 51 can be written as,
\begin{equation}
V^{i}_{12} = M^{i}_{22} + M^{i}_{32} = 0
\end{equation}
, where $M^{i}_{22}$ can be written as a function of $v_Q$, but not $v^{*}_L$, due to its $Z_3$ charge,  
\begin{equation}
\begin{split}
&M^{i}_{22} = h^{}_{16} v^{}_{SM} v'_{Q} \omega
\end{split}
\end{equation}
, and that for $M_{32}$, we have the same pattern as $M_{22}$ but with the breakings of Higgs as $v''_{Q}$ instead of $v'_{Q}$ and $\omega^2$ instead of $\omega$ in the expression.  Having such $M_{22}$ and $M_{32}$ leads us to Equation 48.  

Moving on to the third column in Equation 54, we have,
\begin{equation}
\begin{split}
&M^{i}_{23} = h^{}_{17} v^{}_{SM} v'_{NU} v'_{NU} \omega 
\end{split}
\end{equation}
, and that for $M_{33}$ have a same pattern but with the breakings of Higgs as $v''_{Q}$ instead of $v'_{Q}$ and $\omega^2$ instead of $\omega$.  With that, then we end up with,
\begin{equation}
\begin{split}
2 \hspace{0.3mm} v^{}_{NU} v'^{}_{NU} + v'^{}_{NU} v'^{}_{NU} + 2 \hspace{0.3mm} \omega \hspace{0.3mm} v^{}_{NU} v''^{}_{NU} + \omega \hspace{0.3mm} v''^{}_{NU} v''^{}_{NU} = 0
\end{split}
\end{equation}
.  In the leading order, just like that for the charged lepton sector, we end up with,
\begin{equation}
\begin{split}
v'^{}_{NU} + \omega v''^{}_{NU} = 0
\end{split}
\end{equation}
, which is a same pattern as that for the quark sector.  On the other hand, for the Majorana neutrino sector, we have,
\begin{equation}
U^{+}_{L} M^{SE}_{NU,M} =
\begin{array}{c|ccccc|c}
&& V^{i}_{11} & ... & V^{i}_{13} &&\\
&& ... & V^{i}_{22} & ... &&\\
&& V^{i}_{31} & ... & V^{i}_{33} &&\\
\end{array}
\end{equation}
.  Having non-zero $V^{i}_{31}$ and $V^{i}_{13}$ requires breaking of the flavor-specific scalars.  However, due to all the terms in Equation 42 are in a pair with its conjugate,
the breaking that leads us to have non-diagonal terms cannot be realized.  Since mass matrix in the non-sequential part has a same structure as that in Equation 60, we can realize the structure of the mass matrix by adding the terms in the non-sequential sector.


\section{Phenomenology}	
\label{phenomenology}
In our model, calculations for the three generations of masses in the non-sequential sector can be performed similarly as that in \cite{reference2}, given the non-sequential terms in our model being the leading-order terms.  For that reason, we focus on what we need to do to diagonalize the mass matrix with the presence of the sequential terms and how our model can accommodate the empirical result.   

Note that we use a different convention from the previous sections for parameters such as $M^i$ and $V^i$ with different subscript to distinguish the terms in the sequential and the non-sequential sectors.

We start with the quark sector revisiting some of the previous sections.  When we do a rotations by the unitary matrix in Equation 19, we end up with,
\begin{equation}
M^{RO}_{UQ} = U^{+}_{L} M^{}_{UQ}
\end{equation}
, where the mass matrix can be written as a combination of the sequential and the non-sequential part of matrix as,
\begin{equation}
 M^{}_{UQ} = M^{NS}_{UQ} + M^{SE}_{UQ}
\end{equation}
.  Given the presence of the non-diagonal terms in the sequential part after applying a rotation by $U_L$, we do need to rotate the matrix further to fully diagonalize it.  For that, we introduce two rotations, one on the left- and another one on the right-handed.  Then we have,
\begin{equation}
M^{D}_{UQ} = U^{+}_{LO} U^{+}_{L} M^{}_{UQ} U^{}_{RO}
\end{equation}
, where $LO$ represents left-side rotation and $RO$ represents right-side rotation.  Moving the rotation matrices to the left side, we end up with,
\begin{equation}
U^{}_{LO} M^{D}_{UQ} U^{+}_{RO} = U^{+}_{L} M^{}_{UQ}
\end{equation}
.  We moved the two since applying rotations to the diagonal mass matrix is convenient for us to compare element by element.

For the mass matrices, the non-sequential terms can be written as,
\begin{equation}
M^{NS}_{UQ}
=
\begin{array}{c|ccccc|c}
&& M^{i}_{1} & M^{i}_{2} & M^{i}_{3} &&\\
&& M^{i}_{1} & M^{i}_{2} & M^{i}_{3} &&\\
&& M^{i}_{1} & M^{i}_{2} & M^{i}_{3} &&\\
\end{array}
\end{equation}
, where $i=UQ$, up-side quark sector and where $Z_3$ discrete symmetry is held.  The sequential term medicated to the non-sequential sector can be written as,
\begin{equation}
M^{SE}_{UQ}
=
\begin{array}{c|ccccc|c}
&& ... & ... & ... &&\\
&& M^{SE,i}_{21} & M^{SE,i}_{22} & ... &&\\
&& M^{SE,i}_{31} & M^{SE,i}_{32} & ... &&\\
\end{array}
\end{equation}
.  We express the sequential terms as a function of the non-sequential terms using correctional coefficient.  Introducing four coefficients representing ratio between the non-sequential and the sequential terms, then we have,
\begin{equation}
\begin{split}
M^{SE,i}_{mn} = C^{}_{mn} M^{i}_{n}
\end{split}
\end{equation}
, where $m = 2, 3$ and $n = 1, 2$ for the lower-bottom part of the matrix.  With that, $C_{mn}$ can be written as,
\begin{equation}
\begin{split}
C^{}_{22} \sim  h^{ }_{\raisebox{-0.25pt}{\scriptsize $5$}} \hspace{0.2mm} h^{-1}_{50} \hspace{0.2mm} v'^{}_Q \hspace{0.2mm} \Lambda^{-1}, \hspace{0.5mm} C^{}_{21} \sim 2 \hspace{0.4mm} h^{}_{\raisebox{-0.25pt}{\scriptsize $4$}} \hspace{0.2mm} h^{-1}_{40} \hspace{0.2mm} v^{}_Q \hspace{0.2mm} v'_Q \hspace{0.2mm} \Lambda^{-2}
\end{split}
\end{equation}
, where $v'$ is the difference between v.e.v before and after $Z_3$ breaking, and $h_{40}$ and $h_{50}$ are the Yukawa constant for the non-sequential sector in Equation 10.  For $C_{32}$ and $C_{31}$, we need to replace $v'_Q$ by $v''_Q$ in Equation 68.  

Note that, even with the mass and the scale being specified as,
\begin{equation}
\begin{split}
v_Q \sim 10^2 \hspace{0.3mm} GeV, \hspace{0.5mm} \Lambda \sim 10^6 \hspace{0.3mm} GeV
\end{split}
\end{equation}
, we can adjust $v'_Q$ and $h_{4, 40}$ and $h_{5, 50}$ to accommodate different values of $C_{mn}$.

For our convenience, we introduce the variable $V_i$.  For the elements in the first column of Equation 72, we have,  
\begin{equation}
\begin{split}
&V_1 = 3 + C_{21} + C_{31},
V_2 = \omega^2 C_{21} + \omega \hspace{0.4mm} C_{31},\\
&V_3 = \omega \hspace{0.4mm} C_{21} + \omega^2 C_{31},
\end{split}
\end{equation}
.  For the elements in the dimension-five column,
\begin{equation}
\begin{split}
&V_4 = \omega \hspace{0.4mm} C_{22} + \omega^2 C_{32},
V_5 = 3 + C_{22} + C_{32},\\
&V_6 = \omega^2 C_{22} + \omega \hspace{0.4mm} C_{32}
\end{split}
\end{equation}
.  With that, the matrix on the left side of Equation 64 can be written as,
\begin{equation}
M^{RO}_{Q}
=
\begin{array}{c|ccccc|c}
&& V^{i}_{1} \hspace{0.2mm} M^{i}_{1} & ... & ... &&\\
&& V^{i}_{2} \hspace{0.2mm} M^{i}_{1} & V^{i}_{5} \hspace{0.2mm} M^{i}_{2} & ... &&\\
&& V^{i}_{3} \hspace{0.2mm} M^{i}_{1} & V^{i}_{6} \hspace{0.2mm} M^{i}_{2} & V^{i}_{9} \hspace{0.2mm} M^{i}_{3} &&\\
\end{array}
\end{equation}
, where the non-sequential terms are diagonalized.  Since we do not have breaking in the dimension-four column, note that,
\begin{equation}
\begin{split}
&V^{i}_7 = 0, \hspace{0.5mm} V^{i}_8 = 0, \hspace{0.5mm} V^{i}_9 = 3
\end{split}
\end{equation}
.  As far as the rotation matrix goes, we can write the left-handed rotational matrix as,
\begin{equation}
U^{}_{LO}
=
\begin{array}{c|ccccc|c}
&& A^{i} & D^{i} & G^{i} &&\\
&& B^{i} & E^{i} & H^{i} &&\\
&& C^{i} & F^{i} & I^{i} &&\\
\end{array}
\end{equation}
. 
Note that, given the upper-right hand side of the matrix in Equation 72 having no terms in the leading order, we cannot diagonalize the mass matrix by having the left-side matrix only, but we do need a non-unity right-hand side rotational matrix.  For the quark sector, the minimum structure of the mass matrix in Equation 72 can be realized by having the right-handed matrix as,
\begin{equation}
U^{+}_{RO}
=
\begin{array}{c|ccccc|c}
&& 1 & ... & ... &&\\
&& ... & K^{i} & L^{i} &&\\
&& ... & -L^{i} & K^{i} &&\\
\end{array}
\end{equation}
.  With the diagonalized mass matrix being defined as,
\begin{equation}
M^{D}_{UQ}
=
\begin{array}{c|ccccc|c}
&& M^{d}_{1} & ... & ... &&\\
&& ... & M^{d}_{2} & ... &&\\
&& ... & ... & M^{d}_{3} &&\\
\end{array}
\end{equation}
.  Based on what we have on the first column, the left side of Equation 64 leads us to,  
\begin{equation}
\begin{split}
&V^{i}_1 M^{i}_{1} = A^{i}M^{d}_1K^{i}\\
&V^{i}_2 M^{i}_{1} = B^{i}M^{d}_1K^{i}\\
&V^{i}_3 M^{i}_{1} = C^{i}M^{d}_1K^{i}\\
\end{split}
\end{equation}
.  Taking the ratio of the first two expressions in Equation 76,
\begin{equation}
\begin{split}
\frac{V^2_1}{A^{2}} = V^2_1 + V^2_2 + V^2_3
\end{split}
\end{equation}
, and we can do the same for $B^i$ and $C^i$.  On the same token, for the second column,

\begin{equation}
\begin{split}
&V^{i}_4 M^{i}_{2} = D^{i}M^{d}_2K^{i} - G^{i}M^{d}_3L^{i} = 0,\\
&V^{i}_5 M^{i}_{2} = E^{i}M^{d}_2K^{i} - H^{i}M^{d}_3L^{i},\\
&V^{i}_6 M^{i}_{2} = F^{i}M^{d}_2K^{i} - I^{i}M^{d}_3L^{i}\\
\end{split}
\end{equation}
.  For the third column,
\begin{equation}
\begin{split}
&V^{i}_7 M^{i}_{3} = D^{i}M^{d}_2L^{i} + G^{i}M^{d}_3K^{i} = 0,\\
&V^{i}_8 M^{i}_{3} = E^{i}M^{d}_2L^{i} + H^{i}M^{d}_3K^{i} = 0,\\
&V^{i}_9 M^{i}_{3} = F^{i}M^{d}_2L^{i} + I^{i}M^{d}_3K^{i}\\
\end{split}
\end{equation}
.  The first two expressions in Equation 79 and 80 leads us to
\begin{equation}
\begin{split}
D^{i} \hspace{0.2mm} M^{d}_2 + \hspace{0.2mm} G^{i} \hspace{0.2mm} M^{d}_3 = 0
\end{split}
\end{equation}
.  For the right-handed rotation, we have,
\begin{equation}
\begin{split}
|L^i| + |K^i| = 0
\end{split}
\end{equation}
, which also leads us to,
\begin{equation}
\begin{split}
E^{i} \hspace{0.2mm} M^{d}_2 + \hspace{0.2mm} H^{i} \hspace{0.2mm} M^{d}_3 = 0
\end{split}
\end{equation}
.  From which we can calculate $D$, $G$, $E$ and $H$, and the rest can be calculated based on $U_{LO}$ being an unitary matrix.  

Note that, just like that is shown for the down-side quark sector, we have two correction terms, $C_{22}$ and $C_{21}$ in the up-side quark sector.  Having another two in the down-side with a same structure of the rotational matrix, we can solve,
\begin{equation}
\begin{split}
U^{}_{\raisebox{-0.25pt}{\scriptsize $CKM$}} =
U^{UQ+}_{\raisebox{-0.25pt}{\scriptsize $LO$}}
U^{DQ}_{LO}.
\end{split}
\end{equation}
, for the total four correction terms by comparing four elements in the left- and the right-hand side of the matrix.

As for the charged lepton sector, the breaking of v.e.v cannot be realized due to the sequential structure of the mass matrix.  In other words, we can choose $U_{LO}$, as long as the the rotation by $U_{LO}$ is countered by $U_{RO}$ to maintain the mass matrix being a diagonal.

As for the neutrino sector, the sequential terms of the Majorana part after the rotation, Equation 43, have the same structure as those of the non-sequential terms thus we do not need to apply an additional rotation.  However, we need one for the Dirac neutrino sector.  

Note that, unlike the quark sector, when combining the sequential and the non-sequential part, we do not need to do the rotation by $U_L$, Equation 18, but $U_{LO}$ only, since the non-sequential Dirac mass matrix as shown in Equation 39 is diagonalized.

Given the mass of the light neutrino sector as in \cite{reference3},
\begin{equation}
M_L = -M_{DN} \hspace{0.2mm} M_{MN} \hspace{0.2mm} M_{DS,T}
\end{equation}
where $T$ stands for transpose, we express $M^{DN}$ as a combination of the sequential and the non-sequential mass matrix as,
\begin{equation}
M^{RO}_{DN} = U^{}_{LO} M^{}_{NS,DN} U^{}_{RO} + U^{}_{LO} M^{}_{SE,DN} U^{}_{RO}
\end{equation}
, where $DN$: Dirac neutrino, $NS$: non-sequential, $SE$: sequential.

Starting with the Dirac mass matrix,
\begin{equation}
M^{RO}_{DN}
=
\begin{array}{c|ccccc|c}
&& V^{i}_1 M^{i}_{1} & ... & ... &&\\
&& V^{i}_2 M^{i}_{1} & V^{i}_5 M^{i}_{2} & V^{i}_8 M^{i}_{3} &&\\
&& V^{i}_3 M^{i}_{1} & V^{i}_6 M^{i}_{2} & V^{i}_9 M^{i}_{3} &&\\
\end{array}
\end{equation}
.  We do the same for the quark sector introducing two more coefficients for the third column, 

\begin{equation}
\begin{split}
M^{SE,i}_{mn} = C^{i}_{mn} M^{i}_{n}
\end{split}
\end{equation}
, where $m = 2,3$ and $n = 1,2,3$.

Since $U_L$ is an identity matrix in the Dirac neutrino sector, we have different patterns in the correction terms compared to those in the quark sector.  For the first column,  
\begin{equation}
\begin{split}
&V^{i}_1 = 1, \hspace{0.3mm}
V^{i}_2 = C_{21}, \hspace{0.3mm} 
V^{i}_3 = C_{31} 
\end{split}
\end{equation}
, and for the elements in the second column,
\begin{equation}
\begin{split}
V^{i}_5 = 1 + C_{22}, \hspace{0.3mm}
V^{i}_6 = C_{32}
\end{split}
\end{equation}
, and the third column, 
\begin{equation}
\begin{split}
V^{i}_8 = C_{23}, \hspace{0.3mm} 
V^{i}_9 = 1 + C_{33}
\end{split}
\end{equation}
.  The coefficient for the correction terms can be written as a function of v.e.v and the Yukawa interaction,
\begin{equation}
\begin{split}
&C^{}_{21} \sim  2 \hspace{0.3mm} h^{ }_{\raisebox{-0.3pt}{\scriptsize $15$}} \hspace{0.2mm} h^{-1}_{13} \hspace{0.2mm} v'^{}_{Q} \hspace{0.2mm} \Lambda^{-1},
C^{}_{22} \sim  h^{ }_{\raisebox{-0.3pt}{\scriptsize $16$}} \hspace{0.2mm} h^{-1}_{13} \hspace{0.2mm} v^{}_{Q} \hspace{0.2mm} v'^{}_{Q} \hspace{0.2mm} \Lambda^{-2},
\hspace{0.2mm}
\end{split}
\end{equation}
.  For the third column, due to the breaking of $\phi_{NU}$, we have,
\begin{equation}
\begin{split}
&C_{23} \sim  2 \hspace{0.3mm} h^{ }_{\raisebox{-0.3pt}{\scriptsize $17$}} \hspace{0.2mm} h^{-1}_{13} \hspace{0.2mm} v^{}_{NU} \hspace{0.2mm} v'^{}_{NU} \hspace{0.2mm} \Lambda^{-2} 
\end{split}
\end{equation}
.  For $C_{31,32,33}$, the correction terms in the bottom row of the matrix, we need to replace $v'_{}$ by $v''_{}$ in Equation 92 and 93.  

Note that we can keep the pattern of $U_{RO}$ for the Dirac neutrino to be same as that of the Majorana matrix, thus the result shown in \cite{reference3} as the rotation matrix for the non-sequential terms can be held in the leading order.  

In our case, we can write the rotation as,
\begin{equation}
U^{+}_{RO}
=
\begin{array}{c|ccccc|c}
&& K^{i} & ... & L^{i} &&\\
&& ... & 1 & ... &&\\
&& -L^{i} & ... & K^{i} &&\\
\end{array}
\end{equation}
.  With that, from Equation 85, when we multiply the mass matrix by the rotational matrix, for dimension-six part, we have,
\begin{equation}
\begin{split}
&V^{i}_1 M^{i}_{1} = A^{i}M^{d}_1K^{i_{}} - G^{i}M^{d}_3L^{i}_{},\\
&V^{i}_2 M^{i}_{1} = B^{i}M^{d}_1K^{i}_{} - H^{i}M^{d}_3L^{i}_{},\\
&V^{i}_3 M^{i}_{1} = C^{i}M^{d}_1K^{i}_{} - I^{i}M^{d}_3L^{i}_{}\\
\end{split}
\end{equation}
.  For dimension-five,
\begin{equation}
\begin{split}
&V^{i}_4 M^{i}_{2} = D^{i}M^{d}_2 = 0,\\
&V^{i}_5 M^{i}_{2} = E^{i}M^{d}_2,\\
&V^{i}_6 M^{i}_{2} = F^{i}M^{d}_2
\end{split}
\end{equation}
.  For dimension-four,
\begin{equation}
\begin{split}
&V^{i}_7 M^{i}_{3} = A^{i}M^{d}_1L^{i}_{} + G^{i}M^{d}_3K^{i}_{} = 0,\\
&V^{i}_8 M^{i}_{3} = B^{i}M^{d}_1L^{i}_{} + H^{i}M^{d}_3K^{i}_{},\\
&V^{i}_9 M^{i}_{3} = C^{i}M^{d}_1L^{i}_{}+I^{i}M^{d}_3K^{i}_{}
\end{split}
\end{equation}
.  Taking the ratio of the last two expressions in Equation 96, we end up with,  
\begin{equation}
\begin{split}
\frac{V^2_5}{E^{2}} = V^2_5 + V^2_6, 
\end{split}
\end{equation}
\begin{equation}
\begin{split}
\frac{V^2_6}{F^{2}} = V^2_5 + V^2_6
\end{split}
\end{equation}
.  On the same token, taking the square of the second expressions in Equation 95 and 97, we can calculate $B^i$ and $H^i$.  We can do the same for $C^i$ and $I^i$ from the last expressions in Equation 95 and 97.  With that and the matrix being unitary, we can calculate the size of the elements in $U_{LO}$ for the Dirac neutrino sector.  

Given Equation 86 and the tribimaximal matrix \cite{reference3}, we can diagonalize the mass matrix in the light neutrino sector,  The diagonalized mass matrix can be written as,
\begin{equation}
\begin{split}
M^{}_{L,D} = -M^{}_{DN,D} \hspace{0.2mm} M^{}_{MN} \hspace{0.2mm} M^{}_{DS,T,D}
\end{split}
\end{equation}
, where $D$ stands for diagonalized and,
\begin{equation}
\begin{split}
M^{ }_{\raisebox{-0.3pt}{\scriptsize $DN, D$}} = 
U^{ }_{\raisebox{-0.3pt}{\scriptsize $CTRM$}}
U^{DN}_{LO}
M^{ }_{\raisebox{-0.3pt}{\scriptsize $DN$}}
U^{DN}_{RO}
\end{split}
\end{equation}
, where $CTRM$ stands for corrected tribimaximal matrix,
\begin{equation}
\begin{split}
U^{}_{CTRM} \sim U^{}_{TRM}
\end{split}
\end{equation}
.  It is needed since the diagonalized matrix in the Dirac neutrino sector in Equation 86 does have the size of the elements being different from each other to the presence of the sequential terms.  


Note that, due to the mass matrix in the charged lepton sector being diagonalized, we can parameterize $U_{LO}$ for the neutrino sector first and then we can parameterize that for the charged lepton matrix so that,
\begin{equation}
\begin{split}
U^{ }_{\raisebox{-0.3pt}{\scriptsize $PMNS$}} = U^{+}_{CL} \hspace{0.3mm}
U^{ }_{\raisebox{-0.3pt}{\scriptsize $CTRM$}}
\hspace{0.3mm} U^{DN}_{LO}
\end{split}
\end{equation}
, where $CL$ stands for the charged lepton sector.  Such can be realized in our model since the mass matrix in the charged lepton sector is diagonalized without rotations, thus having a non-identity $U_{CL}$ can be countered by their right-handed matrix to keep the matrix being still diagonalized.

We have three correction terms, $C_{21}$, $C_{22}$ and $C_{23}$, in the Dirac neutrino sector.  With one or more degrees of freedom being introduced from the charged lepton sector, $U_{CL}$, we can construct $U_{PMNS}$ in Equation 103.

\section{Discussion}	
\label{discussion}
Taking the contents of the model in the hierarchical $U(1)$ gauge but building the interaction under $A_4$ rotational symmetry, we could illustrate that the terms in the sequential gauge could be understood as the correction terms in the mass matrix, and that from the rotational symmetry in \cite{reference2} as the leading-order terms, without introducing super-symmetry.  
Under $A_4$, the breaking of the family universality could be a reason for seeing the mixing among the contents.  In other words, the breaking of the family universality could be one of the causes for having the correction terms in the mass matrix.  Furthermore, $Z_3$ is not broken for $\phi_{SM}$ but all the other Higgs scalars in our model thus that leaves a door open for us to incorporate all that was built for the non-sequential models.  

For the future studies, it would be interesting to see how the different types of the seesaw mechanism can be understood when we take the hierarchical gauge as the correctional terms.  In $A_4$, it is type 1 seesaw mechanism where the generation of the mass can be described, but it is mainly type 3 in the sequential gauge scheme.  Accommodation of the two using the approach similar to that shown in our study but with details along with numerical analysis could be one of future studies.  In addition, we can do the study presented here but under different flavor symmetries such as $S_4$ and $D_4$, studying more generalized cases of assigning the $U(1)_X$ sequential charges differently along with discrete $Z_n$ for different generations.  
Investigating the mediation between the charged and neutral lepton sector and their correlation to quark-lepton complementarity \cite{reference16} and imposing additional constraints from the previous studies could be another subject too.  



\section*{Acknowledgments}
The author truly thanks his family for all their support.  

\section*{Appendix}
We list all the fields introduced in our model.  The considered symmetry is,
\begin{equation}
G = SU(2)_L \times U(1)_Y \times U(1)_X \times A_4 {,}
\end{equation}
where $U(1)_X$ is {auxiliary charge} and $A_4$ is a tetrahedron rotational symmetry.
{The field and scalar} contents of the model for the sequential part are given by:
\begin{equation}
\begin{split}
&\psi_{iL} \sim \left[ \hspace{0.4mm} 2, {-1}, 4n, 1_i \hspace{0.4mm} \right],\\
&\phi_{SM} \sim \left[ \hspace{0.4mm} 2, {1}, c_1, 3\hspace{0.4mm} \right]\\
&\phi_{L} \sim \left[ \hspace{0.4mm} 1, 0, c_2, 3\hspace{0.4mm} \right]
,\phi_{Q} \sim \left[ \hspace{0.4mm} 1, 0, c_3, 3\hspace{0.4mm} \right],\\
&l_{i,R} \sim \left[ \hspace{0.4mm} 1, {-2}, 5n, 1_i \hspace{0.4mm} \right]\label{particlecontent}\\
&u_L, d_L \sim \left[ \hspace{0.4mm} 2, 1/3, 0, 3\hspace{0.4mm} \right],\\
&u_{i,R} \sim \left[ \hspace{0.4mm} 1, 4/3, -3n, 1_i \hspace{0.4mm} \right],\\
&d_{i,R} \sim \left[ \hspace{0.4mm} 1, -2/3, 3n, 1_i \hspace{0.4mm} \right]\\
\end{split}
\end{equation}
, where {$1_i= [\hspace{0.4mm} 1,1',1'' \hspace{0.4mm}]$, representing the generation, $1'$ and $1''$ in $A_4$, to indicate the non-trivial and $1$ the trivial representation} respectively.

For non-sequential sector, we have,
\begin{equation}
\begin{split}
&\phi_{NU} \sim \left[ \hspace{0.4mm} 1, 0, c_1, 3\hspace{0.4mm} \right],\\
&\psi^{NS}_{L} \sim \left[ \hspace{0.4mm} 2, -1, 0, 3 \hspace{0.4mm} \right],\\
&{l^{NS}_{i,R}} \sim \left[ \hspace{0.4mm} 1, -2, 0, 1_i \hspace{0.4mm} \right],\\
&u^{NS}_{i,R} \sim \left[ \hspace{0.4mm} 1, 4/3, 0, 1_i\hspace{0.4mm} \right],\\
&d^{NS}_{i,R} \sim \left[ \hspace{0.4mm} 1, -2/3, 0, 1_i\hspace{0.4mm} \right],\\
&v^{SE}_{i,R} \sim \left[ \hspace{0.4mm} 3, 0, n, 1_i\hspace{0.4mm} \right]
\end{split}
\end{equation}
, where the convention for $1_i$ is same as that of the sequential sector.

The $Z_3$ discrete symmetry charge is assigned as,
\begin{equation}
\begin{split}
&\psi^{SE}_{1,L} \sim \omega, \psi^{SE}_{2,L} \sim \omega^2, \psi^{SE}_{3,L} \sim \omega,
v^{SE}_{1,R} \sim \omega^2, v^{SE}_{2,R} \sim \omega, v^{SE}_{3,R} \sim \omega^2,\\
&\phi^{}_{Q} \sim \omega, \phi^{}_{L} \sim \omega, \phi^{}_{NU} \sim \omega,\\
&\psi_{L} \sim \omega, l_{i,R} \sim \omega
\end{split}
\end{equation}
, where $i = 1, 2, 3$.

All the charges in the model can be written as a function of,
\begin{equation}
n = \frac{3}{4}-\frac{i}{4}
\end{equation}
, with $U(1)_X$ charge for the Higgs scalars being assigned as,
\begin{equation}
c_1 = 0, \hspace{0.4mm} c_2 = \frac{1}{4}, \hspace{0.4mm} c_3 = \frac{3}{4}.
\end{equation}


\begin{thebibliography}{99}

\bibitem{reference18}
T. Kobayashi, H. Ohki, H. Okada, Y. Shimizu and M. Tan -imoto, Introduction to non-abelian discrete symmetries for particle physicists, Springer, 2022.

\bibitem{reference17}
I. Aitchison and A. Hey, Gauge theory in particle physics, Fourth edition, CRC Press, 2013.

\bibitem{reference1}
V. Vien, Lepton mass and mixing in a neutrino mass model based on $S_4$ flavor symmetry, Int. J. Mod. Phys. A, 31 165 0039, 2016,

\bibitem{reference2}
V. Vien and H. Long, Neutrino mixing with non-zero $\theta_{13}$ and $CP$ violation in the 3-3-1 model based on $A_4$ flavor sy -mmetry, Int. J. Mod. Phys. A, 30 1550117, 2015.

\bibitem{reference3}
X. He, Y. Keum and R. Volkas, $A_4$ flavour symmetry break -ing scheme for understanding quark and neutrino mixing angles, JHEP, 0604 039, 2006.

\bibitem{reference4}
E. Ma, Hierarchical quark and lepton masses with sequential $U(1)$ gauge symmetry, Phys. Lett, B, 136145, 2021.

\bibitem{reference5}
C. Bouchiat, J. Iliopoulos and P. Meyer, An anomaly free version of Weinberg's model, Phys. Lett. B, 38 519, 1972.

\bibitem{reference6}
V. Vien, $B$-$L$ model with $A_4 \times Z_3 \times Z_4$ symmetry for active-sterile neutrino mixing, J. Phys. G: Nucl. Part. Phys, 49 085001, 2022.

\bibitem{reference7}
E. Ma, A new $U(1)$ gauge symmetry of quarks and leptons, Mod. Phys. Lett. A, 17 535, 2002.

\bibitem{reference8}
S. M. Barr, B. Bednarz and C. Benesh, Anomaly constra -ints and new $U(1)$ gauge bosons Phys. Rev. D 34235, 1986.

\bibitem{reference9}
H. Ishimori et. al., Non-Abelian discrete symmetries in particle physics, Prog. Theor. Phys. Suppl.  183, 2010.

\bibitem{reference10}
G. Altarelli and F. Feruglio, Tri-bimaximal neutrino mixing from discrete symmetry in extra dimension, Nucl. Phys. B 720 64-88 2005.

\bibitem{reference11}
F. Bazzocchi, S. Kaneko and S. Morisi, A susy $A_4$ model for fermion masses and flavor mixing, JHEP 0803, 063 2008.

\bibitem{reference12}
A. Blum, C. Hagedorn and M. Lindner, Fermion masses and mixings from tetrahedron rotational symmetries with preserved subgroups, Phys. Rev. D 77 076004 2008.

\bibitem{reference13}
E. Ma, Pathways to naturally small neutrino masses, Phys. Rev. Lett. 81 1171-1174 1998.

\bibitem{reference14}
Y. Koide, H. Nishiura, K. Matsuda, T. Kikuchi and T. Fukuyama, Universal texture of quark and lepton mass matrices and a discrete symmetry $Z_3$, Phys. Rev. D 66 093006 2002.

\bibitem{reference15}
D. Hernandez and A. Smirnov, Lepton mixing and discrete symmetries, Phys. Rev. D 86 053014 2012.

\bibitem{reference16}
H. Minakata and A. Smirnov, Neutrino mixing and quark lepton complementarity, Phys. Rev. D 70 073009 2004.

\bibitem{reference17}
G. Altarelli and F. Feruglio, Tri-bimaximal neutrino mixing from discrete symmetry in extra dimensions, Nucl. Phys. B 720 64-88 2005.

\bibitem{reference18}
V. Vien, Cobimaximal neutrino mixing in the  extension with  symmetry,  Modern Physics Letters A 35 2020 2050311.

\bibitem{reference19}
V. Vien and H. Long, $A_4$-based model with linear seesaw scheme for lepton mass and mixing, Physica Scripta 98 2023 015301.

\bibitem{reference20}
V. Vien, Neutrino Phenomenology and keV Dark Matter in the Two-Higgs Doublet Model with $A_4$ Symmetry, Progress of Theoretical and Experimental Physics, 2024 103B01.

\bibitem{reference21}
V. Vien, $A_4 \times Z_2 \times Z_4$ flavor symmetry model for neutrino oscillation phenomenology, Revista Mexicana de Fısica 70 2024 060801.

\end{thebibliography}

\end{document}